\documentclass[iop,apjl]{emulateapj}
\usepackage{apjfonts,graphicx}

\slugcomment{}

\shorttitle{The stellar and dynamical masses of GCs}
\shortauthors{Sollima et al.}

\begin{document}


\title{A comparison between the stellar and dynamical masses of six globular
clusters}


\author{A. Sollima\altaffilmark{1}, M. Bellazzini\altaffilmark{2} and J.-W.
Lee\altaffilmark{3}}


\altaffiltext{1}{INAF Osservatorio Astronomico di Padova, vicolo
dell'Osservatorio 5, I-35122 Padova, Italy.}
\altaffiltext{2}{INAF Osservatorio Astronomico di Bologna, via Ranzani 1,
I-40127 Bologna, Italy.}
\altaffiltext{3}{Department of Astronomy and Space Science, Sejong University, Seoul 143-747, Korea}


\begin{abstract}
We present the results of a comprehensive analysis of the structure and
kinematics of six Galactic globular clusters. By comparing the results of the most 
extensive photometric and kinematical surveys available to date with
suited dynamical models, we determine the stellar and dynamical masses of these stellar 
systems taking into account for the effect of mass segregation, anisotropy 
and unresolved binaries. We show that the stellar masses of these clusters are
on average smaller than those predicted by canonical integrated stellar 
evolution models because
of the shallower slope of their mass functions. The derived stellar masses are
found to be also systematically smaller than the dynamical masses by 
$\sim$ 40\%, although the presence of systematics affecting our estimates cannot
be excluded. If confirmed, this evidence can be linked to an increased fraction
of retained dark remnants or to the presence of a modest amount of dark matter.  
\end{abstract}





\section{Introduction}

Among the stellar system zoo there is a widely recognized discrepancy between
the photon-emitting mass (constituted by stars and gas) and dynamical mass (including
the contribution of dark matter). From galaxy clusters to dwarf galaxies the
mass of the baryonic component, estimated through the conversion of
light in mass, has been found to be significantly smaller than that estimated
via kinematical considerations (see Tollerud et al. 2011 for a recent review).
The commonly accepted solution to this problem invokes the presence of a large
amount of dark, non-baryonic matter driving the kinematics of these systems
without contributing to their luminosity. Within this framework, 
globular clusters (GCs) are considered a notable exception: although they are
the next step down from the smallest stellar systems containing dark matter 
(the Ultra Compact Dwarf galaxies; Mieske et al. 2008) and brighter than the
most dark matter dominated systems (the Ultra Faint Galaxies; Simon et al. 2011)
they are not
believed to contain dark matter (Baumgardt et al. 2009). This conclusion is mainly justified by the
evidence that dynamical mass-to-light (M/L) ratios of GCs are 
$\sim$25\% {\it smaller} than those predicted by simple stellar population models 
that assume a canonical Initial Mass Function (IMF; McLaughlin \& van der Marel
2005; Strader et al. 2009, 2011). However, these models do not account for dynamical effects such as the 
preferential loss of low-mass stars due to energy equipartition which alters the
shape of the present-day mass function (PDMF) and consequently the stellar M/L ratio.
In fact, all the past determinations of PDMFs in GCs derived slopes shallower
than that of a canonical (e.g. Kroupa 2001) IMF (Piotto \& Zoccali 1999; Paust et al.
2010; De Marchi et al. 2010). In a recent paper Kruijssen \& Mieske (2009) modeled the evolution of the mass
function of 24 GCs including the effects of dynamical dissolution, low-mass star 
depletion, stellar evolution and stellar remnants finding a substantial agreement
between the prediction of their model and the dynamical M/L ratio estimated
through the cluster kinematics. They conclude that dynamical effects are likely
responsible for the observed discrepancy between stellar and dynamical masses.
The flood of available data from deep photometric and
spectroscopic surveys make now possible to test this issue observationally at an
higher level,
sampling the mass function of GCs down to the hydrogen burning limit (Paust et
al. 2010) and their velocity dispersion profile up to many half-mass radii (Lane
et al. 2010a). This gives the unprecedented opportunity to derive stellar masses
from direct star counts instead of converting light in mass, obtaining an
estimate which does not require the assumption of a $M/L$ ratio.

In this paper we compare the results of deep Hubble Space Telescope (HST) photometric
observations and wide-field radial velocities available from recent public 
surveys with a set of dynamical models with the aim of deriving the amount of 
stellar and dynamical mass in a sample of six Galactic GCs.

\section{Observational material}

For the present analysis we make use of three different databases:
\begin{itemize}
\item{The set of publicy available deep ACS@HST photometric catalogs of the "globular cluster
treasury project" (Sarajedini et al.
2007)\footnote{http://www.astro.ufl.edu/\~ata/public\_hstgc/};}
\item{The radial velocities obtained by the AAOmega@AAT survey of GCs (Lane et al.
2011)\footnote{http://cdsarc.u-strasbg.fr/viz-bin/qcat?J/A+A/530/A31};}
\item{The surface brightness profiles derived by Trager et al. (1995)\footnote{ftp://cdsarc.u-strasbg.fr/pub/cats/J/AJ/109/218}.}
\end{itemize}

The photometric database consists of high-resolution HST observations of a
sample of 65 Galactic GCs. The database has been
constructed using deep images secured with the Advanced Camera for Surveys (ACS)
Wide Field Channel through the F606W and F814W filters. The field of view of the camera
($202\arcsec\times202\arcsec$) is centered on the clusters center with a
dithering pattern to cover the gap between the two chips, allowing a full 
coverage of the core of all the six GCs considered in our analysis.
This survey provides deep color-magnitude diagrams (CMD) reaching the
faint Main Sequence (MS) of the target clusters down to the hydrogen burning limit
(at $M_{V}\leq10.7$) with a signal-to-noise ratio $S/N>10$.
The results of artificial star experiments are also available to allow an
accurate estimate of the completeness level and photometric errors.
A detailed description of the photometric reduction,
astrometry and artificial star experiments can be found in Anderson et al.
(2008). Since all the considered GCs have relaxation times shorter than their
ages (Mc Laughlin \& van der Marel 2005), the effects of mass segregation are expected to be important. It is
therefore useful to constraint the radial variation of the MF outside the
half-mass radius. Unfortunately, the ACS treasury survey covers only the central
part of the cluster at distances $<2\arcmin$ from the cluster centers.
Additional deep data sampling the external part of the cluster are 
available only for NGC288. They consist of
a set of ACS images observed under 
the program GO-12193 which are centered 
at $\sim5\arcmin$ away from the cluster center. 
In particular, $3\times200~s$ + $1\times15~s$ long exposures have been taken 
through the
F606W filter and $3\times150~s$ + $1\times10~s$ long exposures through the
F814W one. All images were passed through the standard ACS/WFC reduction 
pipeline. Data reduction has been performed on the individual pre-reduced (.flt)
images using the SExtractor photometric package (Bertin \& Arnouts 1996). 
For each star we measured the flux contained within a radius of 0.1 arcsec 
(corresponding to 2 pixel $\sim$FWHM) from the star center. 
We preformed the source detection on the stack of all images while the 
photometric analysis has been performed independently on each image. 
Only stars detected in two out of three long-exposures or in the short ones have been included in the 
final catalogue. We used the most isolated and brightest stars in the field 
to link the aperture magnitudes at 0.5 arcsec to the instrumental ones, 
after normalizing for exposure time. Instrumental magnitudes have been 
transformed into the VEGAMAG system by using the photometric zero-points by 
Sirianni et al. (2005). Finally, each ACS pointing has been corrected for 
geometric distortion using the prescriptions by Hack \& Cox (2001).
We preformed artificial star experiments on the science frames following
the prescriptions described in Sollima et al. (2007). In summary, a set of
artificial stars have been simulated using the Tiny Tim model of the ACS PSF 
(Krist et al. 2010) and added to all images at random positions within
$36\times36$px cells centered on a grid of $100\times50$ knots along the x and y
directions of the two ACS chips (a single star for each cell). 
The F814W magnitude of each star has been 
extracted from a luminosity function simulated adopting a Kroupa (2001) mass
function which has been converted in F814W magnitude using the mass-luminosity
relation of Dotter et al. (2007). The corresponding F606W magnitude has been
derived by interpolating along the cluster mean ridge line of the 
F814W-(F606W-F814W) CMD. We performed the photometric reduction on
the simulated frames with the same procedure adopted for the science frames
producing a catalog of $\sim10^{5}$ artificial stars.  

The radial velocity database has been derived by Lane et al. (2009, 2010a,
2010b, 2011) from a large number of spectra of Red Giant Branch (RGB) stars
observed in fields centered on
10 Galactic GCs. Spectra were obtained with the multi-fiber AAOmega spectrograph
mounted at the Anglo Australian Telescope with the 1700D and the 1500V gratings
on the red and blue arm, respectively. With this configuration spectra covering
the Ca II triplet region (8340-8840 \AA) and the swathe of iron and magnesium 
lines around $\sim 5200$\AA~were
obtained with a resolution of R=10000 and R=3700 for the red and blue arms,
respectively. A detailed description of the reduction procedure and radial
velocities estimates can be found in Lane et al. (2009, 2010a, 2010b, 2011).
For the present work we adopted the radial velocities extracted with the RAVE
pipeline since they provide a better estimate of the radial velocity uncertainty
when compared with the available high-resolution spectroscopic studies (see
Bellazzini et al. 2012). In the analysis we used only stars satisfying the same 
membership criterion adopted by these authors and within $d<2~r_{h}$ from the
cluster center (see Sect. \ref{method}); they are 132 in NGC288, 164 in
NGC5024, 190 in NGC6218, 357 in NGC6752, 676 in NGC6809 and 193 in NGC7099.

The surface brightness profiles were taken from the Trager et al. (1995)
database. They were constructed from generally inhomogeneous data based mainly on
the Berkeley Globular Cluster Survey (Djorgovski \& King 1984). The surface
brightness profile of each cluster has been derived by matching several set of
data obtained with different techniques (aperture photometry on CCD images and
photographic plates, photoelectric observations, star counts, etc.). Although
these inhomogeneities may represent a drawback of the analysis (see Sect.
\ref{bias}), 
this database represents the unique collection of wide field surface brightness
profiles covering the entire extent of the target clusters up to their tidal
radius.

The target clusters analyzed in this work were selected among the 10 objects in common between the
three above databases requiring that {\it i)} the photometric observations reach
the magnitude of stars with masses $M\sim0.15~M_{\odot}$ with a completeness
level $\geq50\%$, and {\it ii)} the systemic rotation does not exceed the 10\%
of the cluster velocity dispersion. This last requirement is necessary since
rotation affects the determination of the actual velocity dispersion (and
consequently the dynamical mass) both providing a source of rotational kinetic 
energy and introducing an azimuthal variation of star counts and kinematics
(Wilson 1975) which will not be included in our models. Six clusters satisfy the
above criteria, namely: NGC288, NGC5024, NGC6218, NGC6752, NGC6809 and NGC7099.

\section{Dynamical models}
\label{models}

\begin{figure}
\epsscale{1.2}
\plotone{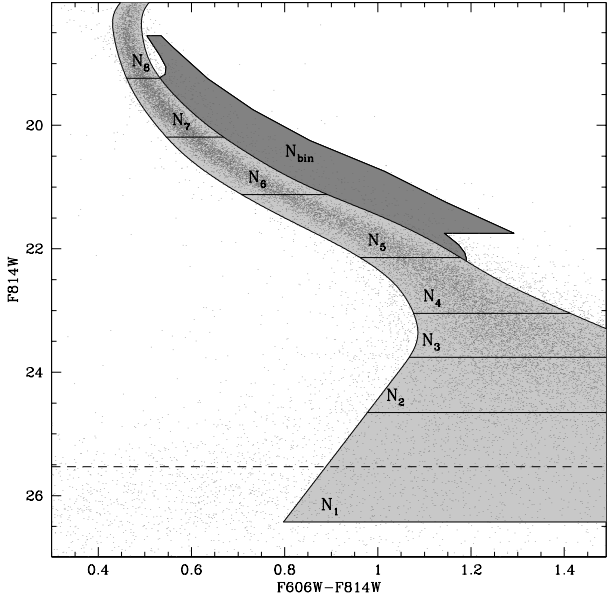}
\caption{Selection boxes adopted for
the population of single stars (from $N_{1}$ to $N_{8}$) and binaries
($N_{bin}$) of NGC~288. The F814W-(F606W-F814W) color-magnitude diagram is overplotted. The
magnitude corresponding to the 50\% completeness is marked by dashed line.}
\label{show}
\end{figure}

Following the prescriptions of Gunn \& Griffin (1979), the phase-space distribution 
of stars is given by the contribution of eight groups of masses with the 
distribution function 
\begin{eqnarray}
\label{eq_df}
f(E,L)&=&\sum_{i=1}^{8} k_{i} exp\left(-\frac{A_{i}L^{2}}{2\sigma_{K}^{2}r_{a}^{2}}\right)
\left[ exp\left(-\frac{A_{i}E}{2\sigma_{K}^{2}}\right)-1 \right]\nonumber\\     
\sum_{i=1}^{8}f_{i}(r,v_{r},v_{t})&=&\sum_{i=1}^{8}k_{i}
exp\left[-A_{i}\frac{v_{t}^{2}}{2\sigma_{K}^{2}}\left(\frac{r}{r_{a}}\right)^{2}\right]~\times\nonumber\\
& &\left[exp\left(-\frac{A_{i}(v_{r}^{2}+v_{t}^{2}+2\psi)}{2\sigma_{K}^{2}}\right)-1
\right]\nonumber\\
\end{eqnarray}
where $E$ and $L$ are, respectively, the energy and angular momentum per unit mass, 
$v_{r}$ and $v_{t}$ are the radial and tangential components of the velocity, 
the effective potential $\psi$ is the difference between the cluster potential 
$\phi$ at a given radius r and the potential at the cluster tidal radius 
$\psi \equiv \phi - \phi_{t}$, $A_{i}$ and $k_{i}$ are scale factors for each mass group, 
and $\sigma_{K}$ is a normalization term. 
For each cluster we defined the eight mass bins (all covering equal mass intervals at 
different ranges) to sample the entire mass
range from the hydrogen burning limit to the mass at the end of the Asymptotic
Giant Branch evolution set by comparison between the CMD and the isochrones by
Dotter et al. (2007)(see Fig. \ref{show}). The dependence on mass of the 
coefficients $A_{i}$ determines the degree of mass segregation of the cluster.
Since the half-mass relaxation times
of all the six GCs of our sample are significantly smaller than their ages
(McLaughlin \& van der Marel 2005) we adopted $A(i)\propto m_{i}$ (Michie 1963).
The parameter $r_{a}$ determines the 
radius at which orbits become more radially biased. As 
$r_{a}\rightarrow \infty$, models become isotropic.
As usual, the above distribution function has been integrated to obtain the 
number density and the radial and tangential components of the velocity dispersion of
each mass group: 
\begin{eqnarray}
\label{eq_struc}
n_{i}(r)&=&4\pi
\int_{0}^{\sqrt{-2\psi}}\int_{0}^{\sqrt{-2\psi-v_{r}^{2}}} v_{t}
f_{i}(r,v_{r},v_{t}) dv_{t}dv_{r},\nonumber\\
\sigma_{r,i}^{2}&=&\frac{4\pi}{\rho_{i}(r)}\int_{0}^{\sqrt{-2\psi}}v_{r}^{2}\int_{0}^{\sqrt{-2\psi-v_{r}^{2}}} v_{t}
f_{i}(r,v_{r},v_{t}) dv_{t}dv_{r},\nonumber\\
\sigma_{t,i}^{2}&=&\frac{4\pi}{\rho_{i}(r)}\int_{0}^{\sqrt{-2\psi}}\int_{0}^{\sqrt{-2\psi-v_{r}^{2}}}
v_{t}^{3} f_{i}(r,v_{r},v_{t}) dv_{t}dv_{r}\nonumber\\
\end{eqnarray}
The above equations can be written in terms of dimensionless quantities by substituting 
\begin{eqnarray*} 
\zeta &=&\frac{v_{r}^2}{2\sigma_{K}^{2}}, \eta =\frac{v_{t}^2}{2\sigma_{K}^{2}},\\ 
\tilde{\rho}_{i}&=&\frac{m_{i} n_{i}}{\sum_{j} m_{j} n_{0,j}}, \tilde{r}=\frac{r}{r_{c}},\\ 
W&=& -\frac{\psi }{\sigma_{K}^{2}}, \tilde{r}_{a}=\frac{r_{a}}{r_{c}},
\end{eqnarray*}
where $\rho_{0}$ is the central cluster density and 
$$r_{c}\equiv \left(\frac{9 \sigma_{K}^{2}}{4 \pi G \rho_{0}}\right)^{1/2}$$
is the core radius (King 1966).
The potential at each radius is determined by the Poisson equation
\begin{equation}
\label{eq_poiss}
\nabla^{2}\psi=4\pi G \rho
\end{equation}
Equations \ref{eq_struc} and \ref{eq_poiss} have been integrated once assumed a
value of the adimensional potential at the center $W_{0}$ outward till the
radius $r_{t}$ at which both density and potential vanish.
The total mass of the cluster $M$ is then given by the sum of the masses of all the
groups
\begin{eqnarray*}
M&=&\sum_{i=1}^{8}m_{i}N_{i}\\
 &=&\frac{9 r_{c} \sigma_{K}^{2}}{G}\sum_{i=1}^{8}
\int_{0}^{r_{t}}\tilde{\rho}_{i}\tilde{r}^{2}d\tilde{r}\\
\end{eqnarray*}
As a last step, the above profiles have been projected on the plane of the sky 
to obtain the surface mass density 
$$\Sigma_{i}(R)=2\int_{R}^{r_{t}}\frac{\rho_{i}r}{\sqrt{r^{2}-R^{2}}}dr$$
and the line-of-sight velocity dispersion
$$\sigma_{v,i}^{2}(R)=\frac{1}{\Sigma_{i}}\int_{R}{r_{t}}\frac{\rho_{i}[2\sigma_{r,i}^{2}(r^{2}-R^{2})+\sigma_{t,i}^{2}R^{2}]}{r\sqrt{r^{2}-R^{2}}}dr$$
In the above models the shape of the density and velocity dispersion profiles 
are completely determined by the parameters ($W_{0},\tilde{r}_{a},N_{i}$) while
their normalization is set by the pair of parameters ($r_{c},M$).

For each choice of the above parameters a large number ($N_{tot}=10^{6}$) of particles
has been randomly extracted from a continuous mass function which reproduces the number
counts in each mass bin and distributed across the cluster according to their
masses by interpolating through the density profiles of the various mass groups.
For each particle the tangential and radial components of its velocity have been
then extracted from the distribution function defined by eq. \ref{eq_df} and
projected in an arbitrary direction \footnote{this is equivalent to randomize the 
direction of the line of sight}.
The masses of the particles have been converted in F814W and Johnson V magnitude adopting the
mass-luminosity relation of Dotter et al. (2007). 
Isochrones has been selected adopting the metallicities by Carretta et al. (2009a)\footnote{The metallicity of
NGC~5024, not included in the Carretta et al. (2009a) sample, has been taken 
from Arellano Ferro et al. (2011).} and their magnitudes have been converted 
from the absolute to the apparent plane adopting the distances and reddening 
by Paust et al. (2010)\footnote{The distance and reddening of
NGC~5024 and NGC~6218 are not included in the Paust et al. (2010) sample, and have 
been taken from the latest version of the Harris catalog (Harris 1996, 2010 
edition).}, while ages have been chosen to match the turn-off 
morphology of each cluster. 
We adopted the reddening coefficients listed by Sirianni et al. (2005). 
The F606W magnitude has been
instead assigned by interpolating on the MS mean ridge line of each cluster in
the F814W-(F606W-F814W) CMD. 
The same procedure has been performed for a fraction $f_{b}$ of binaries and 
$f_{remn}$ of dark remnants whose masses and magnitudes have been assigned following the
prescriptions described in Sect. \ref{bin}. A synthetic Horizontal Branch (HB) has been
simulated for each cluster using the tracks by Dotter et al. (2007), tuning the mean mass and mass dispersion
along the HB to reproduce the observed HB morphology. Photometric errors and completeness
corrections have been included in the following way: to each simulated
particle we associated an 
artificial star within $10\arcsec$ and with input F814W magnitude within 
0.25 mag. If the articial star is recovered its corresponding shifts in color 
and F814W magnitude 
have been then applied to the particle, while the particle has been removed
if the associated artificial star has not been recovered in the output 
artificial star catalog. To account for Galactic field interlopers, we used the Galaxy model of 
Robin et al. (2003). A catalogue covering an area of 1 square degree around 
each cluster center has been retrieved. A subsample of stars has been randomly 
extracted from the entire catalogue scaled to the ACS field of view and the 
V and I Johnson-Cousin magnitudes were converted into the ACS photometric 
system by means of the transformations of Sirianni et al. (2005). The
correction for photometric errors and incompleteness have been applied to this
sample as for cluster stars.
So, at the end of the above procedure according to a given choice of
input parameters a mock cluster with
masses, projected velocities and magnitudes accounting for
observational effects has been produced.

\subsection{Binaries and Dark Remnants}
\label{bin}

The presence of binaries can influence both the 
luminosity and the mass estimate of a cluster. Indeed, while the two components
of the binary system contribute to the total budget of light and mass in the
same way as if they were single stars, their mutual interaction has two important
effects: {\it i)} since the two components are not resolved the binary system
can share the same portion of the CMD with single stars of different mass
(Malkov \& Zinnecker 2001), and
{\it ii)} from a dynamical point of view they behave as a single massive object
so that, as energy equipartition is established, they 
present on average lower velocities and a more concentrated distribution with
respect to single stars. The above characteristics make them 
more resistant to the process of mass-loss via evaporation and tidal shocks. It is therefore important to account for the presence of a
fraction of unresolved binaries to properly model the PDMF of the cluster.
To model the population of binaries we extracted $N_{bin}=2 f_{b} N_{tot}$ 
stars from the Kroupa (2001) IMF in the mass range $0.1<M/M_{\odot}<7$ and 
paired them randomly. From the resulting
sample we extracted a subsample of pairs according to a "chance of pairing"
calculated by imposing that {\it i)} the
distribution of mass ratios f(q) (where $q=m_{2}/m_{1}$ is the ratio between the masses of the
secondary and the primary components of each binary system) in the mass range $1<M/M_{\odot}<7$ results the same
observed by Fisher et al. (2005) in their sample of spectroscopic binaries in the solar
neighborhood\footnote{Although the evolution of the characteristics of
binaries in the solar 
neighborhood is expected to be very different from that in GCs we adopted this
sample since it provides the only available constraint on the
binaries mass ratios distribution.}, and {\it ii)} the distribution of systemic
masses must be the same of single stars in the common mass range. The accepted
pairs were added to a library of binaries while the rejected
ones are "broken" and their components have been used for the next iteration. The
procedure has been repeated until all stars are included in the library. 
Then, all binaries whose primary component has a mass exceeding the maximum
allowed for an evolving cluster star have been removed.
The location and velocity of binaries within the cluster has been assigned 
according to their systemic mass (see Sect. \ref{models}). The relative projected velocity of the 
primary component has been added to the projected velocity of each binary
according to the formula
$$v=\frac{2\pi a_{1} sin i}{P(1-e^{2})^{1/2}}[cos(\theta+\omega)+e
cos\omega]~~~~~~~~~\mbox{McConnachie \& C\^ote (2010)}$$
where $a_{1}$ is the semimajor axis of the primary component, $P$ is the orbital 
period, $e$ is the eccentricity, $i$ is the inclination angle to the
line-of-sight, 
$\theta$ is the phase from the periastron, and $\omega$ is the longitude of the 
periastron. We followed the prescriptions of Duquennoy \& Mayor (1991) for the 
distribution of periods and eccentricities. The semimajor axis has been 
calculated using Kepler's third law: 
$$a_{1}=\frac{1}{1+\frac{m_{1}}{m_{2}}}\left[\frac{P^{2}G(m_{1}+m_{2})}{4\pi
^{2}}\right]^{1/3}$$
where $m_{1}$ and $m_{2}$ are the masses of the primary and secondary
components.We removed all those binaries whose corresponding semimajor axes lie
outside the range $a_{min} < a_{1}+a_{2} < a_{max}$ where $a_{min}$ is linked to the radius of the
secondary component (according to Lee \& Nelson 1988) and $a_{max}$ is set physically 
by the maximum separation beyond which the binary becomes unbound due to 
stellar interactions within the cluster (see eq. 1 of Hills 1984).
The distribution of the angles (i, $\theta$, $\omega$) has been chosen 
according to the corresponding probability distributions 
(${\rm Prob}(i)\propto \sin i; {\rm Prob}(\theta)\propto \dot{\theta }^{-1}; {\rm Prob}(\omega)={\rm constant}$).
The magnitudes of the two components have been then estimated as for single
stars and the overall magnitude has been calculated from the sum of their fluxes
in the different passbands. The photometric errors and completeness correction
have been then applied following the same procedure adopted for single stars 
(see Sect. \ref{models}).

The number of dark remnants (white dwarfs, neutron stars and black holes) has
been estimated adopting for the precursors a Kroupa (2001) IMF normalized to the number of stars in
the most massive stellar bin\footnote{Such a choice is driven by the fact that
massive stars are the most resistant against tidal losses which are expected to
decrease the relative fraction of low-mass stars. On the other hand, stellar 
evolutionary
effects are expected to be negligible since the majority of the stars in this
mass bin are located below the turn-off point.}. We then adopted the 
initial-final mass relation and the retention fractions (for the initial 
velocity kicks) by Kruijssen (2009). We applied an additional correction to account for
the selective removal of low-mass remnant as a result of tidal effects by 
imposing the same PDMF in the mass range in common with single stars.

\section{method}
\label{method}

\begin{figure*}
\epsscale{1.}
\plotone{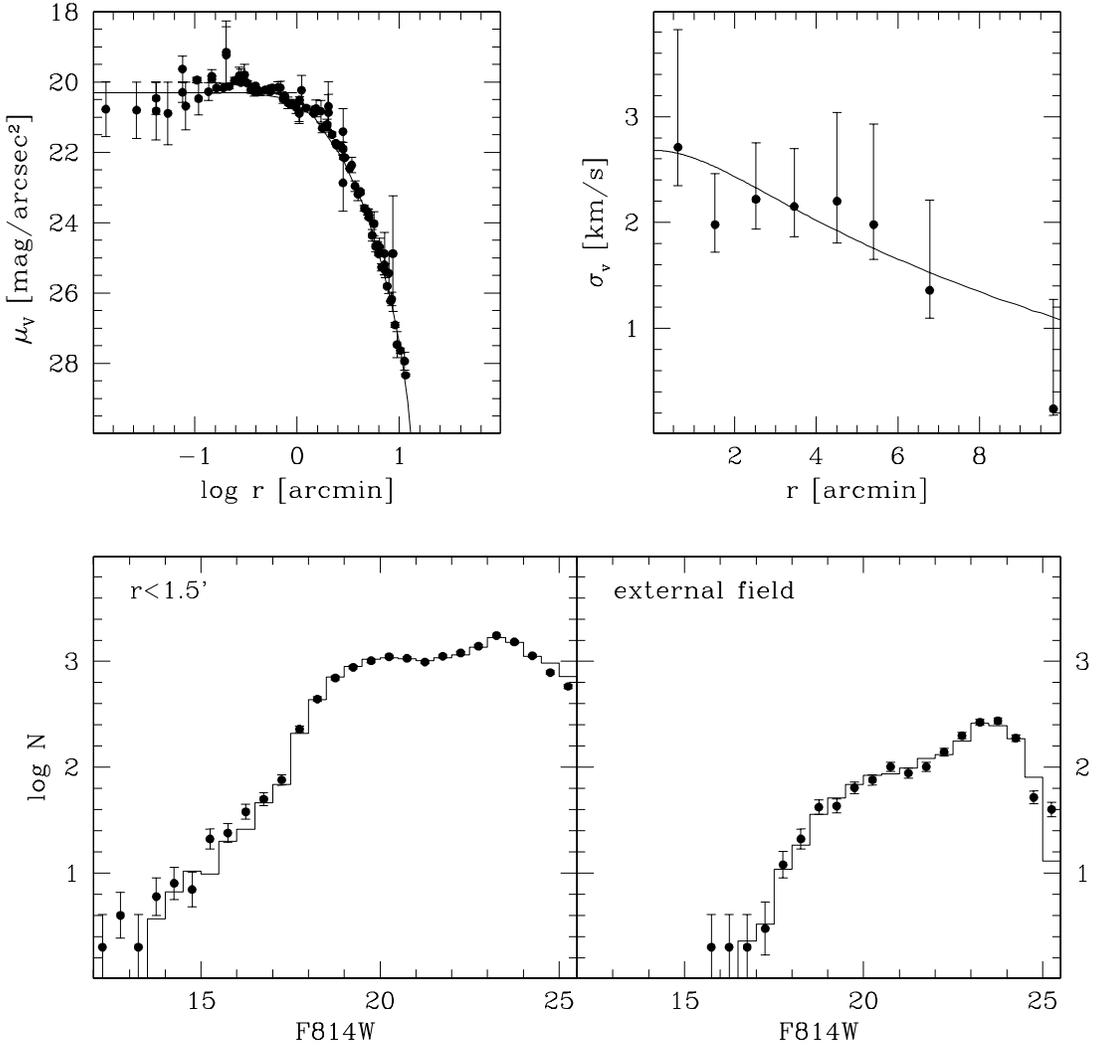}
\caption{Comparison between three observables of NGC~288 (filled points) and the corresponding model 
prediction (solid lines). Upper-left panel: surface brightness profile; upper-right panel:
velocity dispersion profile; bottom panels: F814W luminosity function in the inner (left) and outer 
(right) cluster regions.}
\label{examp}
\end{figure*}

The choice of the best model that reproduces the structure and kinematics of
each cluster has been made by comparing simultaneously four observables 
measured on the databases and in the simulated mock cluster:
\begin{itemize}
\item{The surface brightness profile;}
\item{The F814W luminosity function;}
\item{The fraction of binaries;}
\item{The velocity dispersion profile.}
\end{itemize}

The surface brightness profile of the mock cluster has been derived by summing
the Johnson V fluxes of particles contained within circular concentric annuli 
of variable width and dividing the resulting flux by the annulus area. 
The widths of the annuli have been chosen to contain at least 20 stars to 
minimize Poisson noise while maintaining a good resolution in distance. A
$\chi^{2}$ of the fit has been then calculated.

The comparison of the F814W luminosity function and of the binary fraction have
been performed simultaneously by comparing the number counts ($N_{i}^{obs}$) in nine regions of 
the F814W-(F606W-F814W) CMD defined as follows: eight F814W magnitude intervals 
have been defined in correspondence to the mass bins adopted in the
cluster modeling (see Sect. \ref{models}) and including all stars with colors 
within 3 times the photometric error corresponding 
to their magnitudes. A binary region has been defined within four
boundaries: the selected stars are those contained in magnitude between the 
loci of binaries with primary star mass
$m_{1}=0.45~M_{\odot}$ (faint boundary) and $m_{1}=0.75~M_{\odot}$ (bright
boundary), and in color between the MS ridge line (blue boundary) and the equal-mass
binary sequence (red boundary) both red-shifted by 3 times the photometric 
error (see Fig. \ref{show}; see also Sollima et al. 2007).
We excluded the innermost $0.5\arcmin$ of the core-collapsed clusters NGC6752
and NGC7099 as in these extremely crowded regions the completeness level drops 
below 50\% even at relatively bright magnitudes.  

The velocity dispersion profile of the cluster has been compared using the
likelihood merit function

\begin{equation}
\label{eq_sigma}
l=\sum_{i=1}^{N}(-\frac{(v_{i}-\overline{v})^{2}}{\sigma_{v}(r_{i})^{2}+\delta
v_{i}^{2}}-ln(\sigma_{v}^{2}+\delta
v_{i}^{2}))
\end{equation}

where $r_{i}$, $v_{i}$ and $\delta v_{i}$ are the distance to the cluster
center, the radial velocity and the corresponding uncertainty of
the i-th observed star, $\overline{v}$ is the mean velocity of the sample and $\sigma_{v}(r_{i})$ is the velocity
dispersion predicted for RGB stars by the model at $r_{i}$.
We excluded the outermost region of the cluster at $d>2~r_{h}$ to minimize the
effect of tidal heating that can inflate the velocity dispersion in the
outskirts of our clusters (Johnston et al. 1999; Kupper et al. 2010).

We adopted an iterative algorithm that starts from an initial guess of the
parameters ($W_{0},r_{c},M,\tilde{r}_{a},N_{i},f_{b}$), generates the mock cluster and
calculates the observables described above. Corrections to the parameters are then
calculated using the method described below and the
procedure is repeated until convergence. 
In each iteration the following steps are performed sequentially:
\begin{itemize}
\item{
\begin{enumerate}
\item{The value of $W_{0}$ and $r_{c}$ providing the best fit to the surface
brightness profile are searched by minimizing the $\chi^{2}$ statistic. 
At the end of this step a new mock cluster is generated leaving the 
other parameters unchanged;}
\item{The relative fraction of stars in the eight mass bins $N_{i}$ is adjusted 
by multiplying them for corrective terms which are proportional to the ratio between
the number of objects in each bin of the observed and of the simulated CMDs.
$$N_{i}'=N_{i} \left(\frac{N_{i}^{obs}}{N_{i}^{mock}}
\frac{\sum_{i=1}^{8}N_{i}^{mock}}{\sum_{i=1}^{8}N_{i}^{obs}}\right)^{\eta}$$
where $N_{i}^{obs}$ and $N_{i}^{mock}$ are the number of stars counted in the 
i-th bin in the observed CMD and in simulated one, 
$N'_{i}$ is the updated value of $N_{i}$ and $\eta$ is a softening parameter, 
set to 0.5, used to avoid divergence. An updated value for the binary fraction
$f_{b}$ is also set using the same method.
Then, a new mock cluster is generated
and the procedure is repeated from point {\it (1)} until convergence.}
\end{enumerate}}
\item{The above steps are repeated for different values of $\tilde{r}_{a}$ and
the merit function $l$ is calculated by comparison with the radial velocities to
determine the best-fit value of $\tilde{r}_{a}$.}
\end{itemize}
The procedure usually converges after 10-20 iterations independently on the initial guess
of parameters. Convergence is set when the various parameters start to fluctuate
around the best-fit value with an amplitude (typically $<1\%$) linked to the 
degree of degeneracy among the various solutions. The uncertainty in the derived
parameters has been calculated as the standard
deviations of such fluctuations quadratically combined with the Poisson noise 
of star counts (only for the case of the $M$,$N_{i}$ and $f_{b}$ parameters).
The mass of the model can be constrained in two independent ways: {\it i)} by
best fitting the total number counts (so imposing $\sum_{i=1}^{8}
N_{i}^{mock}=\sum_{i=1}^{8} N_{i}^{obs}$) or {\it ii)} by
matching the amplitude of the velocity dispersion. The former estimate gives the
stellar mass ($M_{*}$), the latter the dynamical one ($M_{dyn}$).

In Fig. \ref{examp} the comparison between the considered observables and the
prediction of the best-fit model are shown for the case of NGC288. It is clear 
that the model well-reproduces the surface brightness and velocity dispersion
profiles as well as the F814W luminosity functions measured in the two radial
samples considered for this cluster. In this cluster, where ancillary
observations are available in an external region, the model that best fit the 
inner sample observables also fits both the shape and the
normalization of the luminosity function in the external sample. Although
this agreement cannot guarantee the full adequacy of models and observables
in the whole sample of clusters, it represents a sanity check supporting the
validity of the assumptions made on mass segregation and on the star
counts-surface brightness conversion.

\section{Results}
\label{res}

\begin{table*}
\caption{Best-fit parameters of the six target clusters}
\begin{tabular}{lccccccccccr}
\tableline\tableline
\tablewidth{0pt}
Name & $W_{0}$ & $r_{c}$ & $r_{c}$   & $\tilde{r}_{a}$ & $f_{b}$ & $f_{b}^{core}$ & $\alpha$ & $log~M_{dyn}$ & $log~M_{*}$
& $M_{dyn}/L_{V}$ & $M_{*}/L_{V}$\\
     &         & $\arcmin$ & pc &                 &   \%    & \%             &       
       & $M_{\odot}$   & $M_{\odot}$ & $M_{\odot}/L_{V,\odot}$ & $M_{\odot}/L_{V,\odot}$\\
\tableline
NGC~288  & 6.2 & 1.90 & 5.40 & 11 & 3.3$\pm$0.2 & 4.8$\pm0.3$ & -1.04$\pm$0.05 & 4.83$\pm$0.07 &
4.73$\pm$0.1& 1.3$\pm$0.4 & 1.0$\pm$0.4\\
NGC~5024 & 11 & 0.55 & 2.99 & +$\infty$ & 4.3$\pm$3.0 & 11.5$\pm$7.2 & -1.08$\pm$0.07 & 5.60$\pm$0.07 &
5.61$\pm$0.1& 1.5$\pm$0.3 & 1.5$\pm$0.4\\
NGC~6218 & 7.7 & 1.05 & 1.50 & 12 & 3.2$\pm$0.2 & 5.8$\pm$0.4 & -0.22$\pm$0.02 & 4.90$\pm$0.05 &
4.74$\pm$0.1& 1.0$\pm$0.3 & 0.7$\pm$0.2\\
NGC~6752 & 13 & 0.42 & 0.51 & +$\infty$ & 0.6$\pm$0.1 & 3.6$\pm$0.6 & -1.41$\pm$0.09 & 5.45$\pm$0.04 &
5.20$\pm$0.1& 2.4$\pm$0.5 & 1.3$\pm$0.4\\
NGC~6809 & 5.5 & 2.50 & 3.32 & 4 & 2.5$\pm$0.2 & 3.4$\pm$0.3 & -0.67$\pm$0.01 & 5.19$\pm$0.03 &
5.08$\pm$0.1& 1.6$\pm$0.2 & 1.3$\pm$0.2\\
NGC~7099 & 13.5 & 0.13 & 0.34 & +$\infty$ & 1.7$\pm$0.1 & 12.0$\pm$0.9 & -1.14$\pm$0.04 & 5.28$\pm$0.05 &
5.08$\pm$0.1& 2.1$\pm$0.4 & 1.3$\pm$0.3\\
\tableline
\end{tabular}
\end{table*}

The procedure described above allows to determine a set of parameters for the
target clusters, in particular: their structural parameters, degree of
anisotropy, binary fraction, PDMF and their stellar and dynamical masses.
The best-fit parameters for each cluster are listed in Table 1. For completeness, 
the $M/L_{V}$ ratios, calculated from the V
magnitudes by van den Bergh et al. (1991), are also listed. The structural
parameter derived here are in good agreement with those estimated by previous
studies (e.g. McLaughlin \& van der Marel 2005). We note that all the clusters
of our sample are best fitted by isotropic (or marginally anisotropic) models.
However, the relatively large uncertainties in the binned velocity dispersion
measures do not allow a meaningful analysis of this issue: the likelihood 
merit function $l$ (see Sect. \ref{method}) vary by less than 10\% over a
relatively large range of $\tilde{r}_{a}$ values in all cases.

The main results on the binary fraction, PDMF and mass are 
discussed in the following sections.

\subsection{Binary fractions}
\label{resbin}

\begin{figure}
\epsscale{1.2}
\plotone{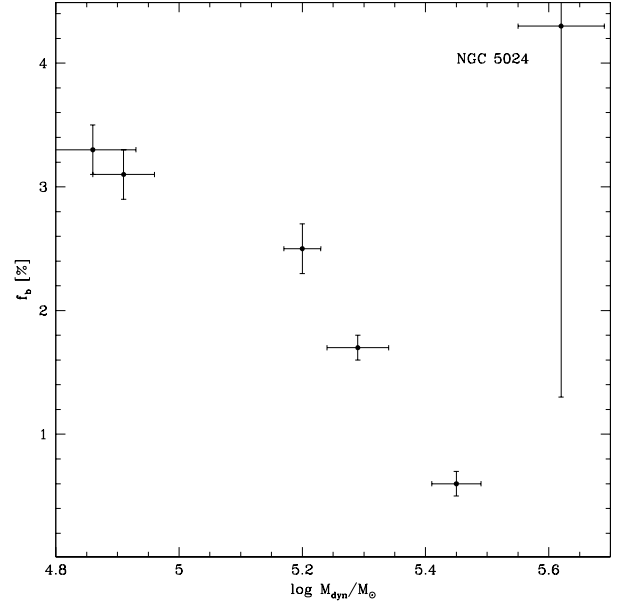}
\caption{Derived fraction of binaries as a function of the dynamical mass.}
\label{binm}
\end{figure}

The binary fractions listed in Table 1 are remarkably lower than
those estimated in previous studies for the clusters in common (Sollima et al. 2007; Milone et al. 2012).
This is related to the distribution of mass
ratios adopted in this work that is different from those used in previous studies. 
The above difference can be well
responsible for the observed discrepancies since in binaries with low mass ratios 
the contribution
of the secondary component to the total flux is negligible, and they appear in
the CMD almost indistinguishable from single stars. Therefore, the adoption of a
mass ratio distribution skewed toward low (high) mass ratios produce a larger
(smaller) correction for hidden binaries.
To verify the consistency of our analysis with the previous works we calculated 
the fraction of binaries in NGC288 adopting a flat mass-ratios distribution over the 
entire mass range instead of the assumptions described in Sect. \ref{bin}. In
this case the overall binary fraction turns out to be $6.5\pm0.3$\% corresponding to
$9.6\pm0.4\%$ in the cluster core, in agreement within the errors with the
estimates by Sollima et al. (2007) and Milone et al. (2012). It is worth noting
that both the PDMF and the mass estimates are almost unaffected by this change:
in this case in fact the slope of the PDMF varies of less than
$\Delta~\alpha\simeq-0.07$ and both the dynamical and luminous masses remain the
same within $\Delta log M/M_{\odot}<0.01$. So, although the adopted mass-ratio 
distribution is somehow arbitrary, it influences only the absolute fraction of 
binaries having a negligible effect on the relative cluster-to-cluster ranking, on the
luminous and dynamical masses and on the shape of the PDMF.

In Fig. \ref{binm} the fraction of binaries of our sample of clusters are plotted
as a function of the derived cluster dynamical mass. Once excluded NGC~5024
which has a large uncertainty in the derived binary fraction, a clear anticorrelation is
noticeable with massive clusters hosting a smaller fraction of binaries. This
result has been already observed in Milone et al. (2008, 2012) and 
Sollima et al. (2010) on a larger sample of clusters.

\subsection{Present-Day Mass Function}

\begin{figure}
\epsscale{1.2}
\plotone{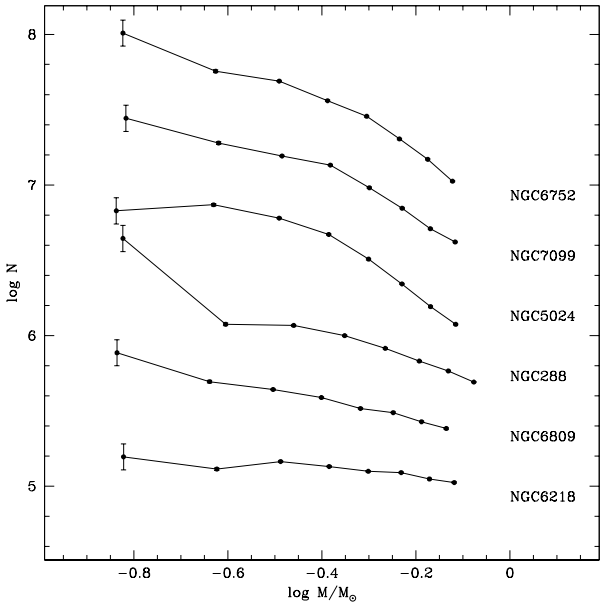}
\caption{The present-day mass function of the six target clusters are shown. A vertical shift has been 
added for each cluster for clarity.}
\label{pdmf}
\end{figure}

In Fig. \ref{pdmf} the global PDMFs of the six clusters are shown. The PDMFs can be well
represented by single power laws whose slopes are all comprised within the
range $-0.2<\alpha<-1.45$ being significantly less steep
than the IMF of Salpeter (1955; $\alpha=-2.35$) and, in all but the case of
NGC6752, of Kroupa (2001; $\alpha=-1.3$ at $M<0.5~M_{\odot}$ with a mean
slope of $<\alpha>=-1.7$ between $0.3<M/M_{\odot}<0.8$). 
The exteme value for NGC~6218 ($\alpha=-0.22\pm0.02$) 
confirms the peculiar flatness of the PDMF of this cluster already noticed by De
Marchi et al. (2006). For the two clusters NGC~288 and NGC7099 in common with 
Paust et al. (2010) the derived slopes agree within the combined
uncertainties of the two works ($\sim 0.2$). In the same way, the slopes derived
for NGC~6809 and NGC~7099 agree to those estimated for these clusters by
Piotto \& Zoccali (1999) and Piotto et al. (1997), respectively.   
We find no signs of a turnover at low masses in all cases,
although a convex shape of the PDMF is visible in NGC~5024. It is however worth
noting that the mass at which the PDMF is expected to culminate is $\sim
0.3~M_{\odot}$ (Paresce \& De Marchi 2000), so this last conclusion depends only on
the relative fraction of stars in the lowest-mass bin. However, the estimate in
this bin is
by far the most uncertain since {\it i)} the mass-luminosity relation in this
mass range is largely uncertain, {\it ii)} the completeness in this bin is low and
uncertain, and {\it iii)} some contamination from faint 
unresolved galaxies falling in the large selection box of this mass bin 
(see Fig. \ref{show}) could be present. So, a firm conclusion on the
presence of a turnover in the PDMF in these clusters cannot be established.

\section{Dynamical versus stellar mass}

\begin{figure*}
\epsscale{1.}
\plotone{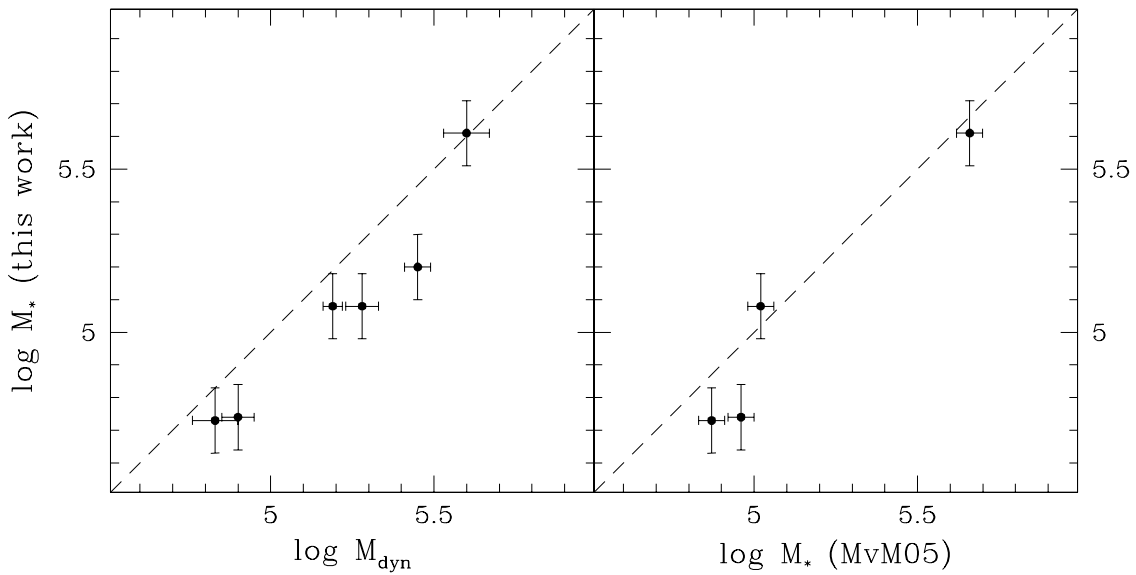}
\caption{Left panel: comparison between the stellar and dynamical masses of the six target
clusters. Right panel: comparison between the stellar masses measured in the
present work and in McLaughlin \& van der Marel (2005). 
The one-to-one relation is marked by the dashed line in both panels.}
\label{dm}
\end{figure*}

\begin{figure*}
\epsscale{1.}
\plotone{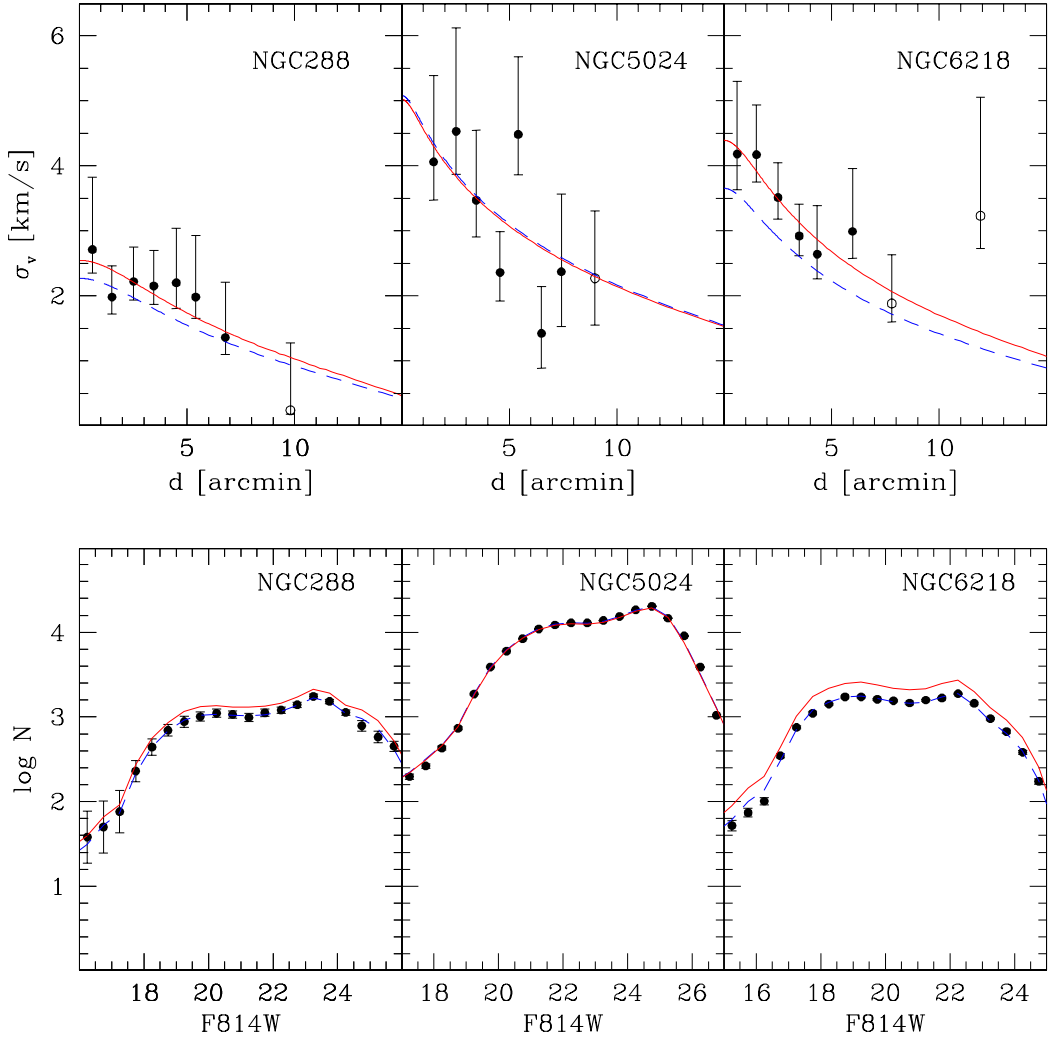}
\caption{Comparison of the observed velocity dispersion profile
 (top panels) and the F814W luminosity function (bottom panels) of the clusters 
 NGC288 (left), NGC5024 (middle) and NGC6218 (right) with the corresponding models 
 with the best-fit $M_{dyn}$ (solid lines; red in the online version) and $M_{*}$
(dashed lines; blue in the online version). The velocity dispersion measures
 not included in the analysis (at $d>2~r_{h}$) are marked with open points in
 upper panels.}
\label{all1}
\end{figure*}

\begin{figure*}
\epsscale{1.}
\plotone{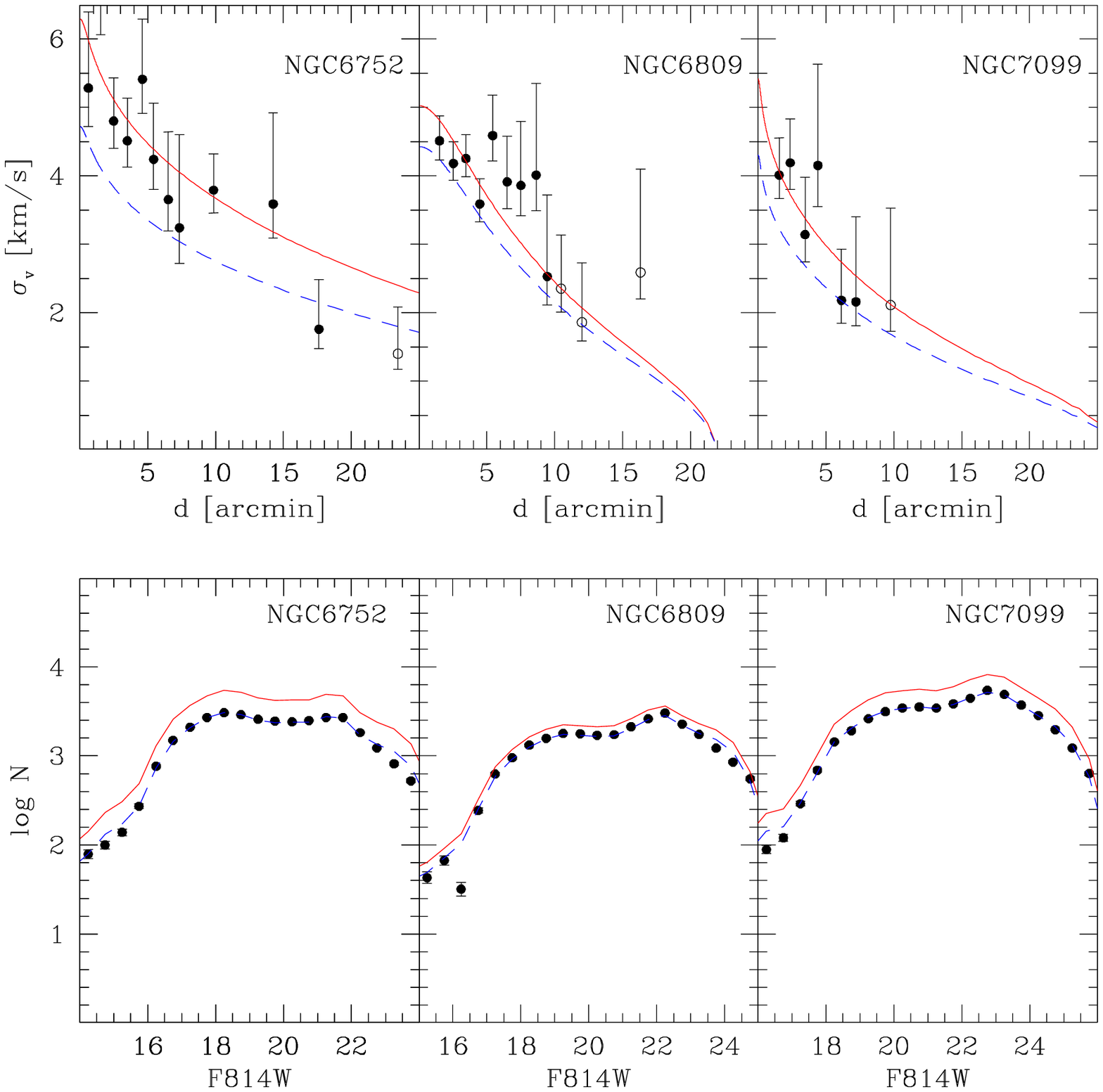}
\caption{Same of Fig. \ref{all1} for NGC6752 (left), NGC6809 (middle) and
NGC7099 (right).}
\label{all2}
\end{figure*}

The main result of this analysis is the measure of the dynamical and
stellar masses of the target clusters. Previous estimates of the dynamical mass
of GCs have been made by Mandushev et al. (1991), Pryor \& Meylan (1993), Lane
et al. (2010a) and Zocchi et al. (2012) which have respectively 3, 5, 6 and 3 clusters in
common with the present study. The mean differences and standard deviations
between the dynamical masses estimated by these authors and those
estimated here are $\Delta~log(M/M_{\odot})=$0.13($\sigma=$0.15),
-0.03(0.17), 0.10(0.26) and -0.09(0.14)\footnote{For the Zocchi
et al. (2012) estimates we adopted those derived with the $f^{(\nu)}$ models,
considered more reliable by these authors. The mean(dispersion) of the
difference in $log~(M/M_{\odot})$ with respect to the estimates made adopting the
King (1966) models is 0.16(0.23).}, respectively, indicating a good
agreement within the errors without significant systematics.
In the right panel of Fig. \ref{dm} the derived stellar masses are compared 
with those reported by McLaughlin \& van der Marel (2005) for the four clusters
in common using the $M/L_{V}$ values predicted by the population synthesis model of Bruzual \&
Charlot (2003) adopting a Kroupa (2001) IMF. The stellar masses
derived here are in reasonable agreement within the errors with those derived by
these authors when considering individual clusters, although on average our
estimates are smaller by $\sim$ 20\%. This is
expected since the Kroupa (2001) IMF is steeper than the PDMFs shown in Fig. 
\ref{pdmf} overestimating the number of low-mass stars that contribute
significantly to the clusters' mass budget and only marginally to their light.

From an inspection of Table 1 it is suddenly evident a discrepancy between the
dynamical and stellar masses of five out of six clusters (with the exception of
NGC~5024) with the dynamical masses systematically larger than the stellar ones 
by $\sim$40\% . This is shown in the left panel of Fig. \ref{dm} where the two 
estimates are compared. 

The observed difference between the dynamical and stellar masses can be
also visualized in Fig. \ref{all1} and \ref{all2} where the velocity dispersion profile
and of the F814W luminosity function of the six target clusters are compared 
with the prediction of the best-fit models with the appropriate $M_{*}$ and
$M_{dyn}$. It is clear that, apart from the case of NGC5024, models are not
able to fit simultaneously both the velocity dispersion profile
and of the F814W luminosity function.
Such a discrepancy is significant
from a statistical point of view: a $\chi^{2}$ test indicates that the
probability that the above difference occurs by chance in the entire sample of
clusters is smaller than 3\%. 
Such a discrepancy can be due to either observational biases in the
determination of masses or to a physical 
reason. In the next sections we will analyze both possibilities.

\subsection{Possible observational biases}
\label{bias}

To understand if the observed deficiency of stellar mass is real or due to an
effect of the adopted technique we must search for some
systematics in our analysis leading to an overestimate of the dynamical masses or
to an underestimate of the stellar ones in all the six clusters of our sample.

The dynamical mass mainly follows from the fit of the velocity dispersion
profile, therefore any systematics affecting the velocity dispersion can alter
its determination. An obvious possibility is that the velocity errors reported
by Lane et al. (2011) are underestimated. In fact, according to 
eq. \ref{eq_sigma}, the observed velocity spread is given by the convolution
of the "true" velocity dispersion and the observational errors. So, for a given
observed velocity spread, the larger are the errors the smaller is the
estimated "true" velocity dispersion and, consequently, the best-fit dynamical mass is 
expected to be smaller than estimated. However, the comparison
with a high-resolution sample shown by Bellazzini et al. (2012) indicates a scatter
compatible with (and even smaller than) the combined uncertainties of the two
datasets. This is also in agreement with what is reported by Lane et al. (2011) who
claim that the velocity uncertainties of their sample could be {\it overestimated}
by $\sim$ 20\%, making the difference even more striking.

The stellar mass depends mainly on the normalization between the observed and
predicted F814W luminosity function and on the correction for the fraction of
mass not covered by the ACS field of view. In all the analyzed clusters the samples of
stars involved in the fit consist of more than 15000 stars, $\sim$90\% of 
them are above the 80\% level of completeness. So, the Poisson error due to number counts and to
the error on the completeness correction are negligible ($\Delta log
M_{*}/M_{\odot}<0.01$).
A good correction for the radial coverage is supposed to be provided by the
excellent fit of the surface brightness profile. However, we must stress that
such profiles are constituted by a compilation of inhomogeneous data, often
merging aperture photometric measurements (tracing the distribution of the brightest
and most massive RGB stars) in the central part with star counts of (low-mass) 
MS stars in the outskirts. As stars with different masses have also different
radial profiles (as a result of mass segregation), this introduces a large
source of uncertainty in the fit. In this context, the simultaneous agreement 
between the predicted and observed F814W luminosity function of NGC~288 in both
the internal and external fields is encouraging.

\subsection{Effect of assumptions}

\begin{figure*}
\epsscale{1.}
\plotone{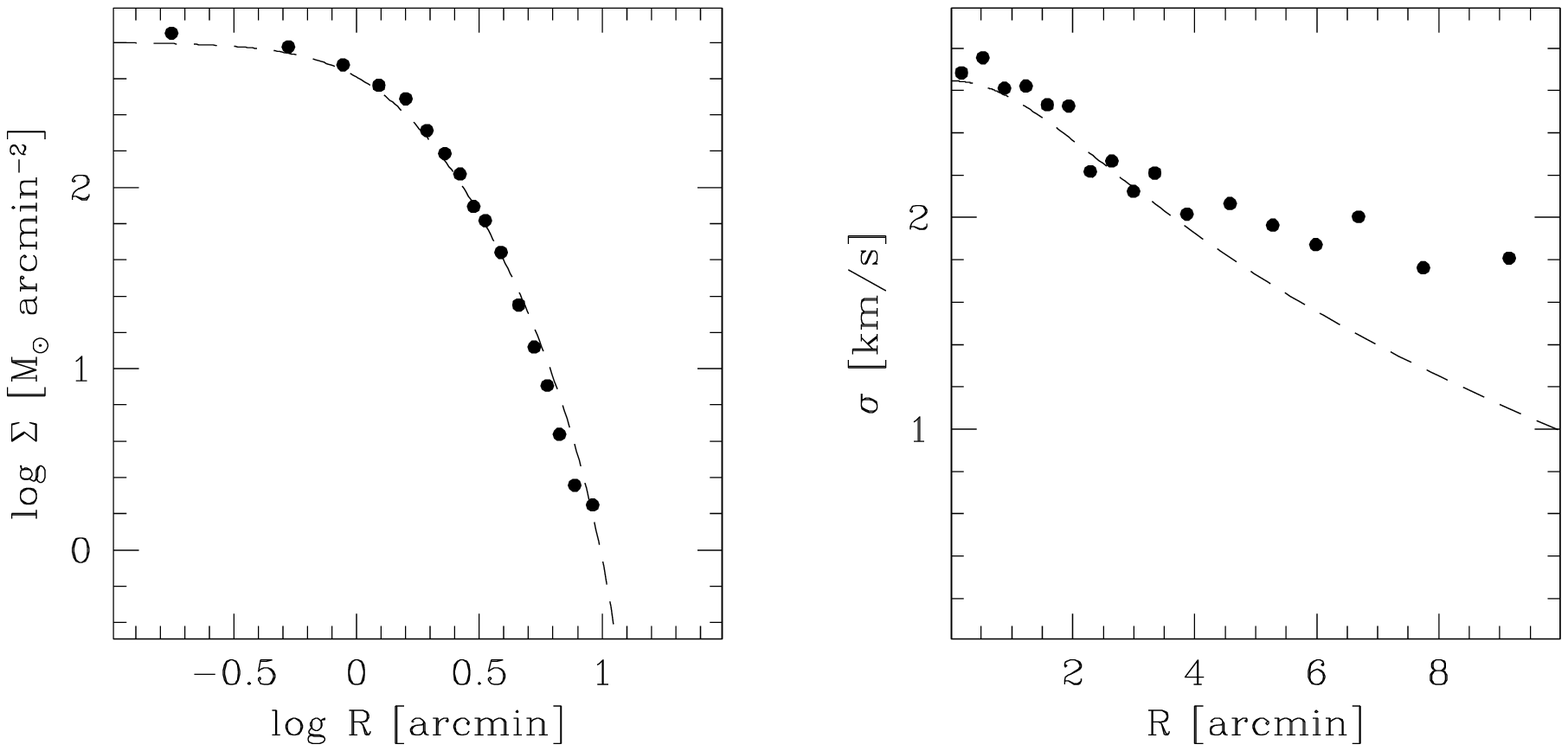}
\caption{Projected density (left panel) and velocity dispersion (right panels) 
profiles 
calculated from the last snapshot of the N-body simulation. The quantities at
the end of the simulation are marked with filled dots. The best-fit static model
is overplotted in both panels with dashed lines.}
\label{nbody}
\end{figure*}

If the discrepancy is not linked to any observational bias, it can be related
to a physical effect not (properly) considered in our models.
There are many possible mechanisms that may affect the mass determination. 
In this Section we evaluate the impact of the various input taking the
cluster NGC288 as tester. In this cluster the observed mass discrepancy ($\Delta
log M/M_{\odot}=0.1$) is smaller than the average of the six analysed
GCs ($\Delta log M/M_{\odot}=0.18$) so that care should be taken in the general
interpretation of the derived results.

As discussed in the previous section, the estimate of the dynamical mass is
sensible to all the parameters affecting the velocity dispersion.
Radial anisotropy can in principle distort the velocity dispersion profile,
resulting in a different normalization of mass. As discussed in Sect. \ref{res},
we take into account for this effect in our models,
but nevertheless the relatively large uncertainties in the velocity estimates do not allow a firm
constraint on this quantity. To test this effect, we repeated the fit imposing
the maximum degree of radial anisotropy that still ensures stability (see Nipoti et al. 
2002). The largest effect on the derived dynamical mass has been found to be
$\Delta~log~M_{dyn}/M_{\odot}\leq 0.03$.
Binaries can also inflate the velocity dispersion because of the presence of the
additional orbital motion of the binary components. Also in this case, while 
our code accounts for such an effect, a different assumption of the distribution
of mass ratios might produce a difference. However, the test performed in Sect.
\ref{resbin} indicates that while the above assumption can affect significantly
the fraction of binaries, the resulting mass is only marginally affected
($\Delta~log~M_{dyn}/M_{\odot}<0.01$).
Another important effect is the heating due to the tidal interaction of the
clusters with the Milky Way: the periodic shocks occurring every disk
crossing and pericentric passage transfer kinetic energy to the cluster stars
inflating its velocity dispersion, particularly in the outskirts (Gnedin \&
Ostriker 1997). While this effect is limited in our case since we
excluded by the fit the outermost portion of the velocity dispersion profile,
Kupper et al. (2010) showed that significant heating can occur also within the
half-mass radius of lose clusters. To test this hypothesis, we run an N-body
simulation to follow the structural and kinematical evolution of a cluster with
the orbital and structural characteristics of NGC~288. This is the less dense
cluster of our sample and should therefore be among those where tidal shocks are
expected to produce the largest effect. We used the last version
of the collisionless N-body code gyrfalcON (Dehnen 2000). The cluster has been represented by 61,440 particles
distributed according to the best-fit multimass model of this
cluster
(see Table 1). The mean masses of the 
particles have been scaled to match the observed cluster mass\footnote{The
adopted scaling does not affect the results of the simulation as the tidal 
effects we are interested to study depends on the overall
cluster potential and not on the individual star masses.}.
We adopted a leapfrog 
scheme with a time step of $\Delta t = 1.45 \times 10^{4} yr$ and a softening length of 
0.43 pc (following the prescription of Dehnen \& Read 2011). Such a relatively 
large time step and softening length do not affect the accuracy of the 
simulation because we are interested in the tidal effects which occur on a
timescale significantly shorter than the relaxation time (the timescale at which
collisions become important). The cluster was launched within the 
three-component (bulge + disk + halo) static Galactic potential of Johnston et 
al. (1995) with the energy and angular momentum listed by Dinescu et al. (1999) 
and its evolution has been followed for 0.852 Gyr (corresponding to 
4 orbital periods). The projected density and velocity dispersion profiles at 
the end of the simulation are shown in Fig. \ref{nbody}. It is clear that 
the final velocity dispersion deviates from the prediction of the steady-state
model at a distance of $\sim4 \arcmin\simeq r_{h}$ to the center. By applying
the same technique described in Sect. \ref{method} to the outcome of the
simulation we overestimate the mass of the cluster by
$\Delta~log~M_{dyn}/M_{\odot}\sim0.06$.

The stellar mass can be underestimated because of an improper conversion
between F814W magnitude and mass or because of the presence of non-luminous
matter. Considering that the mass discrepancy is of the order of $\sim$40\%, 
uncertainties on distance, reddening and age can be excluded as possible causes:
to reproduce the above mass difference one should adopt for these clusters 
ages $<$4 Gyr, distance moduli shifted by $\sim$2.5 mag or reddening
differences of $\Delta E(B-V)\sim$0.2, many orders of magnitudes larger than
their statistical uncertainties. Uncertainties in the mass-luminosity relation
can be also responsible for part of the discrepancy. In this regard, it is worth
noting that while different stellar evolution
models all agree for masses $M>0.1~M_{\odot}$ (Alexander et al. 1997; 
Baraffe et al. 1997; Dotter et al. 2007, Marigo
et al. 2008) significant differences can be present in the least massive bin. 
Also in this case, however, the effect of such uncertainty is not expected to 
exceed few percent in the estimated mass.
A larger effect is instead produced by the
assumption on dark remnants. As described in Sect. \ref{bin}, we followed
the prescription by Kruijssen (2009) for the retention fraction of these 
objects and a Kroupa (2001) IMF for their
precursors. To estimate the extent of the impact of this assumption we 
considered the extreme case of a 100\% retention fraction for all the remnant.
In this ideal case,
the stellar mass would increase by $\Delta~log~M_{*}/M_{\odot}\sim0.08$. 
The adoption of a different slope for the IMF has an even larger effect: the
entire mass discrepancy could be completely removed by assuming a slope of
$\alpha=-1.5$ (instead of -2.3) for masses $M>0.8 M_{\odot}$, because the
shallower the slope of the IMF is, the larger is the relative fraction of massive
stars which terminated their evolution and are now in form of dark remnants.

\section{Discussion}

By comparing the most extensive photometric and spectroscopic
surveys with suitable dynamical models we derived the global binary fraction,
 the PDMF of six Galactic GCs as well as their stellar and dynamical masses.
The approach adopted here has the advantage to not require an a priori
assumption of the $M/L$ ratio to estimate the stellar mass (as usually done in
previous works; e.g. McLaughlin \& van der Marel 2005) which is instead
estimated via direct star counts in the CMD.

We confirm the anticorrelation between binary fraction and cluster mass 
already found in previous studies (Milone et al. 2008, 2012; Sollima et al.
2010) and explained as the consequence of the increased efficiency of the binary 
ionization process in massive clusters.
Indeed, in massive clusters both the number of collisions and the mean kinetic
energy of stellar encounters is larger resulting in a higher probability of
ionization of multiple systems (Sollima 2008).

The PDMFs of the six clusters are well-represented by single power laws, 
although we cannot exclude a turnover at masses $M<0.2~M_{\odot}$, because of the 
uncertainty in the relative fraction of the least massive stars.

The stellar masses derived for these clusters are on average smaller than
those predicted by population synthesis models. This is a consequence of the
fact that the IMF adopted by these models is steeper that the actual PDMF of 
the considered clusters. This has been previously
predicted by Kruijssen \& Mieske (2009) as the result of the preferential loss of
low-mass stars during the cluster evolution.
 
We found a discrepancy between the stellar and dynamical masses of five clusters. 
In particular, dynamical masses are on average $\sim$40\% larger than stellar
ones. Such a discrepancy is statistically significant when considering the
entire sample of six clusters and could not be detected by previous studies
because of the overestimate of the stellar mass mentioned above. 
Unfortunately, we cannot exclude that such a discrepancy is due to the presence
of systematics affecting our estimates. In particular, a possible source of
systematics is constituted by the inhomogeneity of the surface brightness profiles
sample of Trager et al. (1995).
However, if the detected difference would be real, a number of physical
interpretations could be given. After a careful test on the effect of various 
assumptions and physical processes we found that a significant spurious 
increase of the estimated dynamical masses is given by the tidal heating which 
can reduce (but not eliminate) the observed discrepancy.
On the other hand, the assumption on the distribution and retention of dark remnant has
the largest impact on the estimated stellar masses. Part (up to 75\%) of the
observed mass difference can be indeed explained assuming that most the remnants,
including the most massive neutron stars and black holes, are retained. 
This
could be the case if the cluster mass at the moment of the SNe II explosion was
large enough to trap the remnants, whose kinetic energy has been increased by 
the velocity kick following the explosion. The mass necessary to retain the majority of massive remnants 
is of $\sim10^{7}~M_{\odot}$ (see Fig. 1 of Kruijssen 2009). 
Such a large mass is comparable with the initial mass predicted by D'Ercole et 
al. (2008) to ignite the self-enrichment process observed in most GCs (see
Carretta et al. 2009b). Moreover, massive remants are expected to be more
resistant to tidal stripping since tend to sink toward in the cluster core as a
result of mass segregation.
It is however worth noting that a retention fraction larger than 80\% would 
require initial masses and an early mass loss history quite extreme, which has
indeed been proposed in theoretical works to be the result of severe tidal 
shocking in the primordial environment of GCs (Elmegreen 2010; Kruijssen et 
al. 2012).
An even larger effect is provided by the slope of the high-mass end of the IMF 
of the precursors which can
drastically increase the amount of non-luminous mass. In this regard, a recent
analysis by Marks et al. (2012) suggests that the slope of the IMF in the range
$M>M_{\odot}$ could be
actually shallower than the canonical Salpeter (1955) and Kroupa (2001) values
with a dependence on density and metallicity. In this case,
a slope $\alpha \simeq -1.5$ (instead of $\alpha=-2.3$ by Kroupa 2001) could account for the entire difference between the
estimated stellar and dynamical masses. Such a variation would have strong
implications on the evolution of the GC MF: in fact, while the largest mass
contribution would be due to the increased number of white dwarfs, the
consistent increase of the fraction of massive remnants 
(black holes and neutron stars) makes these objects the main drivers of the MF
evolution. It is not clear if such IMF could actually evolve in the PDMF observed in these clusters.

Another possibility is that GCs contain a modest fraction of non-baryonic dark 
matter. In this case, it is possible that what we see now is the remnant of a
larger halo lost during the interaction with the Milky Way (see Saitoh et al.
2006). Some amount of dark matter is also expected if some GCs
are the remnants of past accretion events, as previously suggested by Freeman \&
Bland-Hawthorn (2002). However, given the involved uncertainties, all these 
hypotheses are merely speculative. Further studies will be needed to 
verify the above possibilities.

\acknowledgments

Based on observations made with the NASA/ESA Hubble Space 
Telescope, which is operated by the association of Universities for Research in 
Astronomy, Inc. under the NASA contract NAS 5-26555, under programs GO-10775 
(PI: Sarajedini) and GO-12193 (PI: Lee).
We thank the anonymous referee for his/her helpful comments and suggestions.
A.S. acknowledge the support of INAF through the 2010
postdoctoral fellowship grant.
M.B. acknowledge the financial support of INAF through
the PRIN-INAF 2009 grant assigned to the project Formation
and evolution of massive star clusters, P.I.: R. Gratton.
J.-W. L. acknowledges financial support from the Basic Science Research
Program (grant no.\ 2010-0024954) and the Center for Galaxy Evolution Research
through the National Research Foundation of Korea.


{\it Facilities:} \facility{HST (ACS)}, \facility{AAT (AAOmega)}.




\clearpage

\clearpage



\clearpage



\begin{thebibliography}{}
\bibitem[Alexander et 
al.(1997)]{1997A&A...317...90A} Alexander, D.~R., Brocato, E., Cassisi, S., et al.\ 1997, \aap, 317, 90 
\bibitem[Anderson et al.(2008)]{2008AJ....135.2055A} Anderson, J., 
Sarajedini, A., Bedin, L.~R., et al.\ 2008, \aj, 135, 2055 
\bibitem[Arellano Ferro et al.(2011)]{2011MNRAS.416.2265A} Arellano Ferro, 
A., Figuera Jaimes, R., Giridhar, S., et al.\ 2011, \mnras, 416, 2265 
\bibitem[Baraffe et 
al.(1997)]{1997A&A...327.1054B} Baraffe, I., Chabrier, G., Allard, F., \& Hauschildt, P.~H.\ 1997, \aap, 327, 1054 
\bibitem[Baumgardt et al.(2009)]{2009MNRAS.396.2051B} Baumgardt, H., 
C{\^o}t{\'e}, P., Hilker, M., et al.\ 2009, \mnras, 396, 2051 
\bibitem[Bellazzini et 
al.(2012)]{2012A&A...538A..18B} Bellazzini, M., Bragaglia, A., Carretta, E., et al.\ 2012, \aap, 538, A18 
\bibitem[Bertin 
\& Arnouts(1996)]{1996A&AS..117..393B} Bertin, E., \& Arnouts, S.\ 1996, \aaps, 117, 393 
\bibitem[Bruzual 
\& Charlot(2003)]{2003MNRAS.344.1000B} Bruzual, G., \& Charlot, S.\ 2003, \mnras, 344, 1000 
\bibitem[Carretta et 
al.(2009)]{2009A&A...508..695C} Carretta, E., Bragaglia, A., Gratton, R.,
D'Orazi, V., \& Lucatello, S.\ 2009a, \aap, 508, 695 
\bibitem[Carretta et 
al.(2009)]{2009A&A...505..139C} Carretta, E., Bragaglia, A., Gratton, R., \&
Lucatello, S.\ 2009b, \aap, 505, 139 
\bibitem[D'Ercole et al.(2008)]{2008MNRAS.391..825D} D'Ercole, A., 
Vesperini, E., D'Antona, F., McMillan, S.~L.~W., 
\& Recchi, S.\ 2008, \mnras, 391, 825 
\bibitem[De Marchi et al.(2010)]{2010ApJ...718..105D} De Marchi, G., 
Paresce, F., \& Portegies Zwart, S.\ 2010, \apj, 718, 105 
\bibitem[de Marchi et 
al.(2006)]{2006A&A...449..161D} De Marchi, G., Pulone, L., \& Paresce, F.\ 2006, \aap, 449, 161 
\bibitem[Dehnen(2000)]{2000ApJ...536L..39D} Dehnen, W.\ 2000, \apjl, 536, 
L39 
\bibitem[Dehnen 
\& Read(2011)]{2011EPJP..126...55D} Dehnen, W., \& Read, J.~I.\ 2011, European Physical Journal Plus, 126, 55 
\bibitem[Dinescu et al.(1999)]{1999AJ....117.1792D} Dinescu, D.~I., Girard, 
T.~M., \& van Altena, W.~F.\ 1999, \aj, 117, 1792 
\bibitem[Djorgovski 
\& King(1984)]{1984ApJ...277L..49D} Djorgovski, S., \& King, I.~R.\ 1984, \apjl, 277, L49 
\bibitem[Dotter et al.(2007)]{2007AJ....134..376D} Dotter, A., Chaboyer, 
B., Jevremovi{\'c}, D., et al.\ 2007, \aj, 134, 376 
\bibitem[Duquennoy 
\& Mayor(1991)]{1991A&A...248..485D} Duquennoy, A., \& Mayor, M.\ 1991, \aap, 248, 485 
\bibitem[Elmegreen(2010)]{2010ApJ...712L.184E} Elmegreen, B.~G.\ 2010, 
\apjl, 712, L184 
\bibitem[Fisher et al.(2005)]{2005MNRAS.361..495F} Fisher, J., 
Schr{\"o}der, K.-P., \& Smith, R.~C.\ 2005, \mnras, 361, 495 
\bibitem[Freeman 
\& Bland-Hawthorn(2002)]{2002ARA&A..40..487F} Freeman, K., \& Bland-Hawthorn, J.\ 2002, \araa, 40, 487 
\bibitem[Gnedin 
\& Ostriker(1997)]{1997ApJ...474..223G} Gnedin, O.~Y., \& Ostriker, J.~P.\ 1997, \apj, 474, 223 
\bibitem[Gunn 
\& Griffin(1979)]{1979AJ.....84..752G} Gunn, J.~E., \& Griffin, R.~F.\ 1979, \aj, 84, 752 
\bibitem[Hack 
\& Cox(2001)]{2001acs..rept....8H} Hack, W., \& Cox, C.\ 2001, Instrument Science Report ACS 2001-008, 10 pages, 8 
\bibitem[Harris(1996)]{1996AJ....112.1487H} Harris, W.~E.\ 1996, \aj, 112, 
1487 
\bibitem[Hills(1984)]{1984AJ.....89.1811H} Hills, J.~G.\ 1984, \aj, 89, 
1811 
\bibitem[Johnston et al.(1999)]{1999MNRAS.302..771J} Johnston, K.~V., 
Sigurdsson, S., \& Hernquist, L.\ 1999, \mnras, 302, 771 
\bibitem[Johnston et al.(1995)]{1995ApJ...451..598J} Johnston, K.~V., 
Spergel, D.~N., \& Hernquist, L.\ 1995, \apj, 451, 598 
\bibitem[King(1966)]{1966AJ.....71...64K} King, I.~R.\ 1966, \aj, 71, 64 
\bibitem[Krist et al.(2010)]{2010ascl.soft10057K} Krist, J., Hook, R., 
\& Stoehr, F.\ 2010, Astrophysics Source Code Library, record ascl:1010.057, 10057 
\bibitem[Kroupa(2001)]{2001MNRAS.322..231K} Kroupa, P.\ 2001, \mnras, 322, 
231 
\bibitem[Kruijssen(2009)]{2009A&A...507.1409K} Kruijssen, J.~M.~D.\ 2009, \aap, 507, 1409 
\bibitem[Kruijssen 
\& Mieske(2009)]{2009A&A...500..785K} Kruijssen, J.~M.~D., \& Mieske, S.\ 2009, \aap, 500, 785 
\bibitem[Kruijssen et al.(2012)]{2012MNRAS.421.1927K} Kruijssen, J.~M.~D., 
Pelupessy, F.~I., Lamers, H.~J.~G.~L.~M., et al.\ 2012, \mnras, 421, 1927 
\bibitem[K{\"u}pper et al.(2010)]{2010MNRAS.407.2241K} K{\"u}pper, 
A.~H.~W., Kroupa, P., Baumgardt, H., 
\& Heggie, D.~C.\ 2010, \mnras, 407, 2241 
\bibitem[Lane et al.(2009)]{2009MNRAS.400..917L} Lane, R.~R., Kiss, L.~L., 
Lewis, G.~F., et al.\ 2009, \mnras, 400, 917 
\bibitem[Lane et al.(2010)]{2010MNRAS.401.2521L} Lane, R.~R., Kiss, L.~L., 
Lewis, G.~F., et al.\ 2010b, \mnras, 401, 2521 
\bibitem[Lane et al.(2010)]{2010MNRAS.406.2732L} Lane, R.~R., Kiss, L.~L., 
Lewis, G.~F., et al.\ 2010a, \mnras, 406, 2732 
\bibitem[Lane et 
al.(2011)]{2011A&A...530A..31L} Lane, R.~R., Kiss, L.~L., Lewis, G.~F., et al.\ 2011, \aap, 530, A31 
\bibitem[Lee 
\& Nelson(1988)]{1988ApJ...334..688L} Lee, H.~M., \& Nelson, L.~A.\ 1988, \apj, 334, 688 
\bibitem[Malkov 
\& Zinnecker(2001)]{2001MNRAS.321..149M} Malkov, O., \& Zinnecker, H.\ 2001, \mnras, 321, 149 
\bibitem[Mandushev et 
al.(1991)]{1991A&A...252...94M} Mandushev, G., Staneva, A., \& Spasova, N.\ 1991, \aap, 252, 94 
\bibitem[Marigo et 
al.(2008)]{2008A&A...482..883M} Marigo, P., Girardi, L., Bressan, A., et al.\ 2008, \aap, 482, 883 
\bibitem[Marks et al.(2012)]{2012arXiv1202.4755M} Marks, M., Kroupa, P., 
Dabringhausen, J., \& Pawlowski, M.~S.\ 2012, \mnras, 422, 2246
\bibitem[McConnachie 
\& C{\^o}t{\'e}(2010)]{2010ApJ...722L.209M} McConnachie, A.~W., \& C{\^o}t{\'e}, P.\ 2010, \apjl, 722, L209 
\bibitem[McLaughlin 
\& van der Marel(2005)]{2005ApJS..161..304M} McLaughlin, D.~E., \& van der Marel, R.~P.\ 2005, \apjs, 161, 304 
\bibitem[Michie(1963)]{1963MNRAS.125..127M} Michie, R.~W.\ 1963, \mnras, 
125, 127 
\bibitem[Mieske et 
al.(2008)]{2008A&A...487..921M} Mieske, S., Hilker, M., Jord{\'a}n, A., et al.\ 2008, \aap, 487, 921 
\bibitem[Milone et al.(2008)]{2008MmSAI..79..623M} Milone, A.~P., Piotto, 
G., Bedin, L.~R., \& Sarajedini, A.\ 2008, \memsai, 79, 623 
\bibitem[Milone et 
al.(2012)]{2012A&A...540A..16M} Milone, A.~P., Piotto, G., Bedin, L.~R., et al.\ 2012, \aap, 540, A16 
\bibitem[Nipoti et al.(2002)]{2002MNRAS.332..901N} Nipoti, C., Londrillo, 
P., \& Ciotti, L.\ 2002, \mnras, 332, 901 
\bibitem[Paresce 
\& De Marchi(2000)]{2000ApJ...534..870P} Paresce, F., \& De Marchi, G.\ 2000, \apj, 534, 870 
\bibitem[Paust et al.(2010)]{2010AJ....139..476P} Paust, N.~E.~Q., Reid, 
I.~N., Piotto, G., et al.\ 2010, \aj, 139, 476 
\bibitem[Piotto 
\& Zoccali(1999)]{1999A&A...345..485P} Piotto, G., \& Zoccali, M.\ 1999, \aap, 345, 485 
\bibitem[Piotto et al.(1997)]{1997AJ....113.1345P} Piotto, G., Cool, A.~M., 
\& King, I.~R.\ 1997, \aj, 113, 1345 
\bibitem[Pryor 
\& Meylan(1993)]{1993ASPC...50..357P} Pryor, C., \& Meylan, G.\ 1993, Structure and Dynamics of Globular Clusters, 50, 357 
\bibitem[Robin et 
al.(2003)]{2003A&A...409..523R} Robin, A.~C., Reyl{\'e}, C., Derri{\`e}re, S., \& Picaud, S.\ 2003, \aap, 409, 523 
\bibitem[Saitoh et al.(2006)]{2006ApJ...640...22S} Saitoh, T.~R., Koda, J., 
Okamoto, T., Wada, K., \& Habe, A.\ 2006, \apj, 640, 22 
\bibitem[Salpeter(1955)]{1955ApJ...121..161S} Salpeter, E.~E.\ 1955, \apj, 
121, 161 
\bibitem[Sarajedini et al.(2007)]{2007AJ....133.1658S} Sarajedini, A., 
Bedin, L.~R., Chaboyer, B., et al.\ 2007, \aj, 133, 1658 
\bibitem[Simon et al.(2011)]{2011ApJ...733...46S} Simon, J.~D., Geha, M., 
Minor, Q.~E., et al.\ 2011, \apj, 733, 46 
\bibitem[Sirianni et al.(2005)]{2005PASP..117.1049S} Sirianni, M., Jee, 
M.~J., Ben{\'{\i}}tez, N., et al.\ 2005, \pasp, 117, 1049 
\bibitem[Sollima(2008)]{2008MNRAS.388..307S} Sollima, A.\ 2008, \mnras, 
388, 307 
\bibitem[Sollima et al.(2010)]{2010MNRAS.401..577S} Sollima, A., 
Carballo-Bello, J.~A., Beccari, G., et al.\ 2010, \mnras, 401, 577 
\bibitem[Sollima et al.(2007)]{2007MNRAS.380..781S} Sollima, A., Beccari, 
G., Ferraro, F.~R., Fusi Pecci, F., 
\& Sarajedini, A.\ 2007, \mnras, 380, 781 
\bibitem[Strader et al.(2011)]{2011AJ....142....8S} Strader, J., Caldwell, 
N., \& Seth, A.~C.\ 2011, \aj, 142, 8 
\bibitem[Strader et al.(2009)]{2009AJ....138..547S} Strader, J., Smith, 
G.~H., Larsen, S., Brodie, J.~P., \& Huchra, J.~P.\ 2009, \aj, 138, 547 
\bibitem[Tollerud et al.(2011)]{2011ApJ...726..108T} Tollerud, E.~J., 
Bullock, J.~S., Graves, G.~J., \& Wolf, J.\ 2011, \apj, 726, 108 
\bibitem[Trager et al.(1995)]{1995AJ....109..218T} Trager, S.~C., King, 
I.~R., \& Djorgovski, S.\ 1995, \aj, 109, 218 
\bibitem[van den Bergh et al.(1991)]{1991ApJ...375..594V} van den Bergh, 
S., Morbey, C., \& Pazder, J.\ 1991, \apj, 375, 594 
\bibitem[Wilson(1975)]{1975AJ.....80..175W} Wilson, C.~P.\ 1975, \aj, 80, 
175 
\bibitem[Zocchi et 
al.(2012)]{2012A&A...539A..65Z} Zocchi, A., Bertin, G., \& Varri, A.~L.\ 2012, \aap, 539, A65 

\end{thebibliography}
\end{document}